\documentclass[%
 reprint,
 amsmath,amssymb,
 aps,
prl,
]{revtex4-2}

\usepackage{graphicx}
\usepackage{dcolumn}
\usepackage{bm}
\usepackage{xcolor}
\usepackage[normalem]{ulem}
\usepackage{textcomp}
\usepackage[english]{babel}
\usepackage{natbib}
\begin{document}

\preprint{APS/123-QED}

\title{Single-Shot Characterization of High Transformer Ratio Wakefields in Nonlinear Plasma Acceleration}

\author{R. Roussel}
\email{roussel@ucla.edu}
\author{G. Andonian}
\author{W. Lynn}
\author{K. Sanwalka}
\author{R. Robles}
\author{C. Hansel}
\author{A. Deng}
\author{G. Lawler}
\author{J.B. Rosenzweig}
\affiliation{Department of Physics and Astronomy, University of California, Los Angeles, California 90095, USA}
\author{G. Ha}
\author{J. Seok}
\author{J. G. Power}
\author{M. Conde}
\author{E. Wisniewski}
\author{D. S. Doran}
\author{C. E. Whiteford}
\affiliation{Argonne National Laboratory, Argonne, Illinois 60439, USA}

\date{\today}

\begin{abstract}
Plasma wakefields can enable very high accelerating gradients for frontier high energy particle accelerators, in excess of 10 GeV/m. To overcome limits on total acceleration achievable, specially shaped drive beams can be used in both linear and nonlinear plasma wakefield accelerators (PWFA), to increase the transformer ratio, implying that the drive beam deceleration is minimized relative to acceleration obtained in the wake. In this Letter, we report the results of a nonlinear PWFA, high transformer ratio experiment using high-charge, longitudinally asymmetric drive beams in a plasma cell. An emittance exchange process is used to generate variable drive current profiles, in conjunction with a long (multiple plasma wavelength) witness beam. The witness beam is energy-modulated by the wakefield, yielding a response that contains detailed spectral information in a single-shot measurement. Using these methods, we generate a variety of beam profiles and characterize the wakefields, directly observing beam-loaded transformer ratios up to  $\mathcal{R}=7.8$. Furthermore, a spectrally-based reconstruction technique, validated  by 3D particle-in-cell simulations, is introduced to obtain the drive beam current profile from the decelerating wake data.

\end{abstract}

\maketitle


The development of beam-based plasma wakefield acceleration (PWFA) schemes has provided a potential path to extremely compact high energy accelerators, due to its ability to sustain accelerating gradients well in excess of 10 GeV/m \cite{blumenfeld_energy_2007}.
PWFA uses a highly energetic, high charge bunched beam known as the drive, to excite a nonlinear plasma density wave. The fields in this wave are used to accelerate a second bunch, known as the witness, placed collinearly behind the drive.
While this provides a substantial enhancement in peak accelerating field over current state-of-the-art acceleration schemes, improvements to energy transfer from the drive bunch to the witness, using bunch shaping, are necessary for applications.

Energy transfer from the drive beam to the witness is parameterized by the  transformer ratio, which is defined as $\mathcal{R}=|W_+|/|W_-|$, the ratio between the maximum accelerating field experienced by the witness $|W_+|$ and the maximum decelerating field found inside the drive $|W_-|$. In the context of linear wakefields, it has been shown \cite{bane_collinear_1985,bane_wake_1985} that the limit $\mathcal{R}<$ 2 exists for longitudinally symmetric beams, and that asymmetric beams, such as a linear ramp profile, can exceed this limit. Further work \cite{lemery_tailored_2015} showed that this idea can be extended by modifying the linear ramp in the beam head region to reach even higher transformer ratios.


Methods of producing linearly ramped beam profiles include the use of: self-wakefields in dielectric structures \cite{andonian_generation_2017}; anisochronous dogleg beamlines with nonlinear correction elements \cite{england_generation_2008}; laser pulse stacking at the photocathode; \cite{loisch_photocathode_2018} and emittance exchange methods \cite{ha_precision_2017}. Emittance exchange (EEX) is a process \cite{piot_generation_2011} by which a transverse beam distribution is mapped onto the longitudinal coordinate, a technique which allows flexible longitudinal bunch shaping. EEX may also be used to provide a long, temporally synchronized, witness bunch for single-shot, multi-period wakefield measurements \cite{gao_single-shot_2018}.

Recent methods have been used to perform detailed study of wakefields with high transformer ratios in collinear accelerator schemes including dielectric structures \cite{gao_observation_2018} and plasmas \cite{loisch_observation_2018}.
In the latter, wakefield measurements were made using shaped drive bunches and a short witness bunch, showing high transformer ratios.
However, the laser pulse stacking method of shape generation has an inherent limit in bunch charge achievable, due to space-charge driven degradation of the profile at low energy.
In this Letter we describe experimental measurements that apply the EEX beam shaping technique at moderate energy, mitigating space-charge effects, and providing new tools for wakefield measurements. These beams are employed to characterize plasma wakefield responses from various high-charge, ramped drive bunches, using a single-shot wakefield mapping technique \cite{gao_single-shot_2018} that captures essential elements of the PWFA process and provides insight into the wakefield dependence on bunch shape.

\begin{figure*}[!tbh]
    \includegraphics*[width=\linewidth]{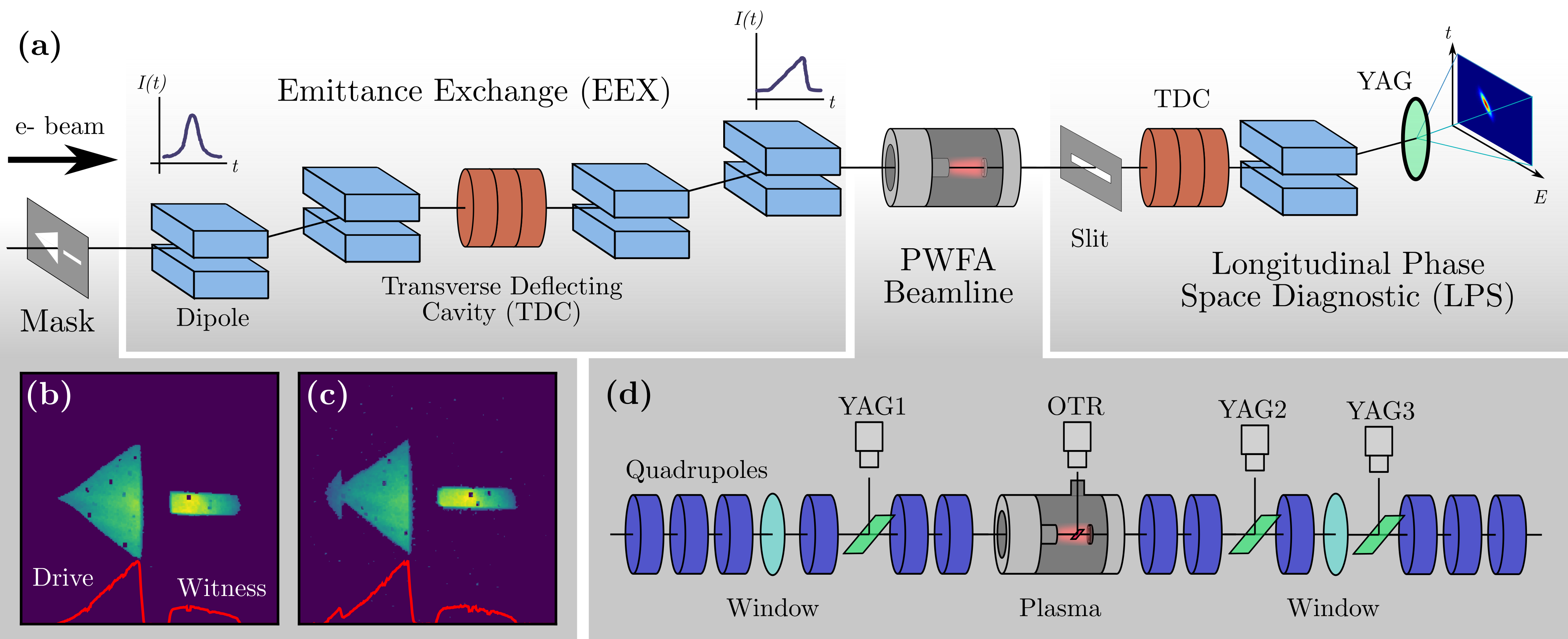}
    \caption{(Color online) Schematic of the beamline at the Argonne Wakefield Accelerator.
    (a) Beams are delivered to the experimental section consisting of a mask, emittance exchange beamline, PWFA beamline and longitudinal phase space diagnostic.
    Two transverse masks were used to shape the horizontal beam projection into a (b) linearly ramped or (c) linear ramp with parabolic head, both with long witness bunches to sample the resulting wakefield.
    (d) The PWFA beamline is shown with major elements including YAG and optical transition radiation (OTR) screens to view the transverse beam profile.}
    \label{fig:1}
\end{figure*}

Experiments were conducted at the Argonne Wakefield Accelerator (AWA) at Argonne National Laboratory. A cesium telluride photcathode based, L-band electron gun produced 15 nC bunches with a pulse length of 6 ps and a near uniform transverse distribution \cite{halavanau_spatial_2017}.
Normal conducting L-band accelerating cavities were then used to accelerate these bunches to 40 MeV. These beams were delivered to the experimental beamline seen in Fig. \ref{fig:1}. A transverse mask was used to shape the horizontal projection of the beam, which is subsequently mapped into the longitudinal coordinate with the EEX beamline.

The EEX beamline (Fig. \ref{fig:1}a) consists of two dogleg sections separated by a transverse deflecting cavity, and is capable of generating a wide range of longitudinal current distributions \cite{ha_precision_2017}.
In the EEX process, a beam particle with a horizontal phase space coordinate of (x$_i$,x$_i^\prime$) is mapped to a final longitudinal position $z_f= (A + B S_x)x_i$ in the linear approximation \cite{piot_generation_2011}, where coefficients A, B are functions defining the EEX beam optics, which include transverse dispersion, longitudinal compression and transverse deflection. The term $S_x = dx'/dx$ represents the slope of the $x-x^\prime$ correlation, which can be tuned via quadrupole settings before the EEX mask to optimize the longitudinal profile \cite{ha_perturbation-minimized_2016}.
Masks were used to shape the horizontal projection of the beam (Fig. \ref{fig:1} b, c), generating two drive profiles, a profile ramped over two plasma wavelengths (Fig. \ref{fig:1}b) and the same profile but with a parabolic head section added (Fig. \ref{fig:1}c).
Both masks included a narrow slit used to create a long, low charge witness bunch placed after the shaped drive.
After the EEX beamline, the longitudinal profile still has a strong correlation with the vertical dimension, which in turn smooths the current distribution after transport through the PWFA beamline (Fig. \ref{fig:1}d) \cite{roussel_transformer_2019}. This correlation also prevents accurate measurement of the beam current in the longitudinal phase space diagnostic, due to horizontal slitting.  

A hollow cathode arc (HCA) \cite{delcroix_hollow_1974} plasma source was used to produce a 6 cm long plasma column with a diameter of 8 mm and a averaged plasma density tunable over the range $n_0=0.3 - 1.3$ x $10^{14}$  cm$^{-3}$ \cite{roussel_externally_2019}.
Argon gas was injected between two concentric tantulum tubes which were heated to 2000 K, creating a thermal electron current high enough to reach the discharge arc regime without the need for a kV class discharge starter.
A 150 V, 200 $\mu$s pulse was used to discharge the plasma up to a current of 300 A. The plasma density is tuned using this current as well as an external solenoid employed to confine the plasma electrons.
The longitudinal plasma density profile on-axis was characterized using a Langmuir triple  probe \cite{chen_instantaneous_1965} which allows a simultaneous, time resolved measurement of the plasma temperature and density.
Beryllium vacuum windows with a thickness of 125 \textmu m were used to isolate the plasma source from the rest of the accelerator, which required an additional beam waist location on either side of the plasma source to reduce emittance growth due to multiple scattering in the windows \cite{lynch_approximations_1991}.  

To characterize the wakefield interaction a longitudinal phase space (LPS) diagnostic consisting of a transverse deflecting cavity and a 20$^\circ$ dipole spectrometer was used.
The transverse deflecting cavity streaks beam particles in the vertical direction according to their arrival time, while the spectrometer dipole maps the energy distribution onto the horizontal dimension.
A horizontal slit, placed directly upstream of the LPS diagnostic, was used to vertically collimate the beam to improve the temporal measurement resolution while reducing spurious energy gain signals from transverse electron motion.

A simulation study informed by beam and plasma measurements was performed using the 3D particle-in-cell (PIC) code WARP \cite{grote_warp_2005}.
The beam size at the interaction point was constrained by optical transmission radiation (OTR) spot size measurements \cite{wartski_interference_1975} and consideration of plasma focusing properties due to a non-adiabatic plasma density ramp \cite{barov_propagation_1994,ariniello_transverse_2019}.
In the simulations, the estimated beam emittance was obtained by using measurements of beam size at the Be input window, along with a calculation of rms divergence due to multiple scattering \cite{lynch_approximations_1991}.
A 40 MeV linearly-ramped drive beam with a length of 5.4 mm, having short fall length, transverse size $\sigma_x = 200$ \textmu m, normalized emittance of 500 mm-mrad and charge of 1.6 nC was injected into a $n_0=1.5\times 10^{14}$ cm$^{-3}$ plasma.
A 4 mm long witness with the same transverse characteristics as the drive and 0.18 nC of charge was injected afterwards to sample the multi-period wake.

\begin{figure}[!tbh]
    \includegraphics[width=\linewidth]{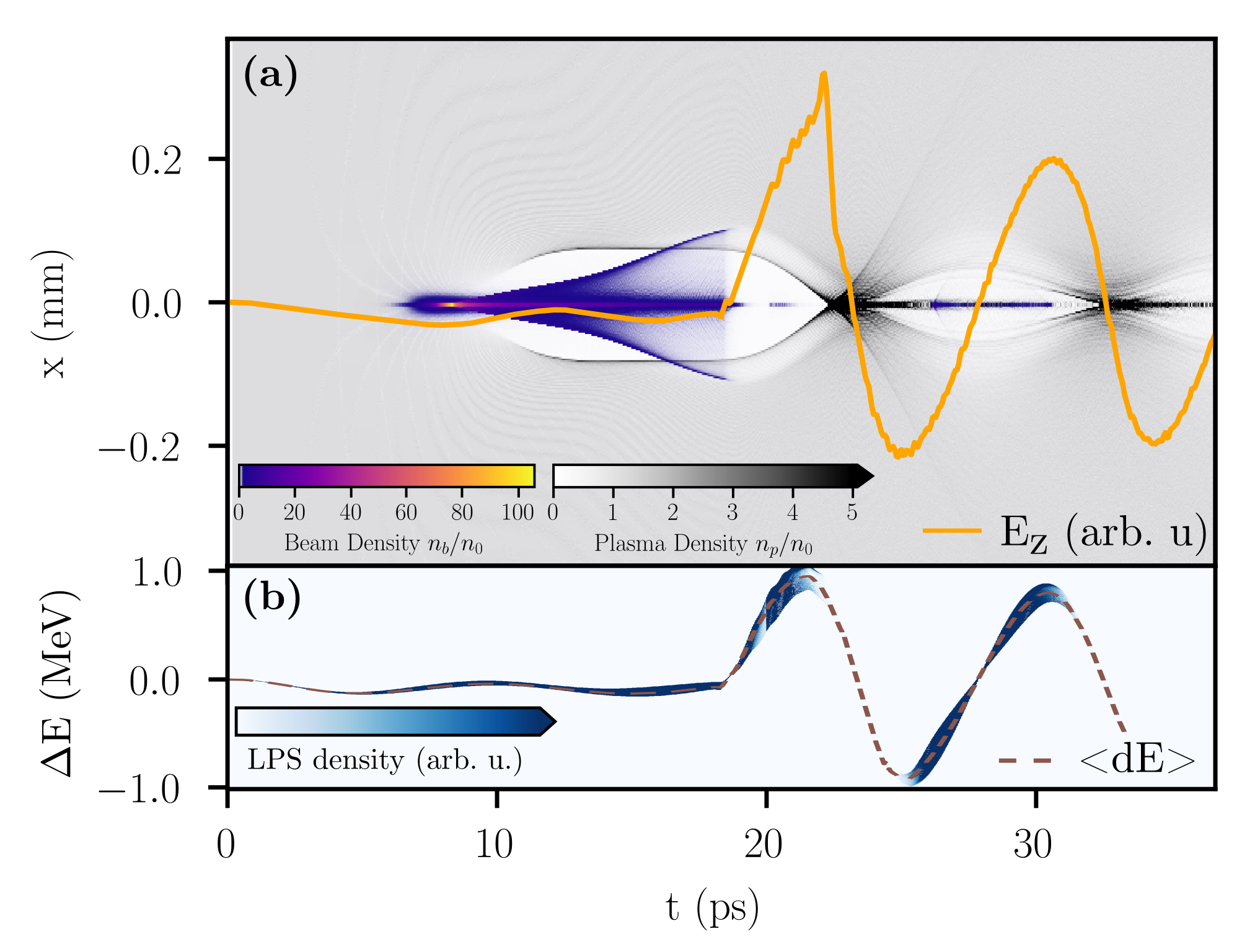}
    \caption{(Color online) Simulation of a linearly ramped drive beam and multi-wavelength witness beam in a $n_0 = $ 1.5 x $10^{14}$ cm$^{-3}$ plasma. 
    (a) Transverse cross section of plasma and beam densities with on-axis longitudinal wakefield (orange).
    (b) Longitudinal phase space of the drive and witness beam (density plot) and time slice energy centroid (brown) for near-axis particles ($|x| <$ 100 $\mu m$.)}
    \label{fig:2}
\end{figure}

The results of the simulation after the beam has traveled 28 mm into the plasma are shown in Fig. \ref{fig:2}.
A clear rarefaction (blowout) region in the plasma is observed, with a density spike occurring afterward ($n_p/n_0 \approx 147$) in Fig. \ref{fig:2}(a). 
The rarefaction boundary is notably off-axis for some distance, collapsing only after the drive current drops to near zero.
The beam head expands freely, followed by a transverse pinch point where the ion column starts to form and then a region where uniform focusing fields from the ion column dominate \cite{barov_propagation_1994}. 
Electrons in the main body of the beam may escape the rarefaction region due to an initial focusing mismatch.

Due to the drives transverse profile evolution, and the resulting change in the wakefield, it is appropriate to examine the time averaged wakefields' contribution to the beam energy change for each temporal slice of the beam rather than assuming an instantaneous quasi-static approximation for the wakefield. In this interpretation, the longitudinal phase space of near-axis particles seen in Fig. \ref{fig:2}(b) reveals the effective transformer ratio over the entire beam-plasma interaction.
Only near-axis electrons are considered as those far off-axis do not take part in the usable interaction, and this condition closely matches the experimental scenario, as these do not pass through the thin horizontal slit in the LPS diagnostic. Calculation of the time-slice energy centroid of the drive and witness yields an averaged wakefield profile having a transformer ratio of 7.3.

\begin{figure}[!tbh]
	\includegraphics[width=\linewidth]{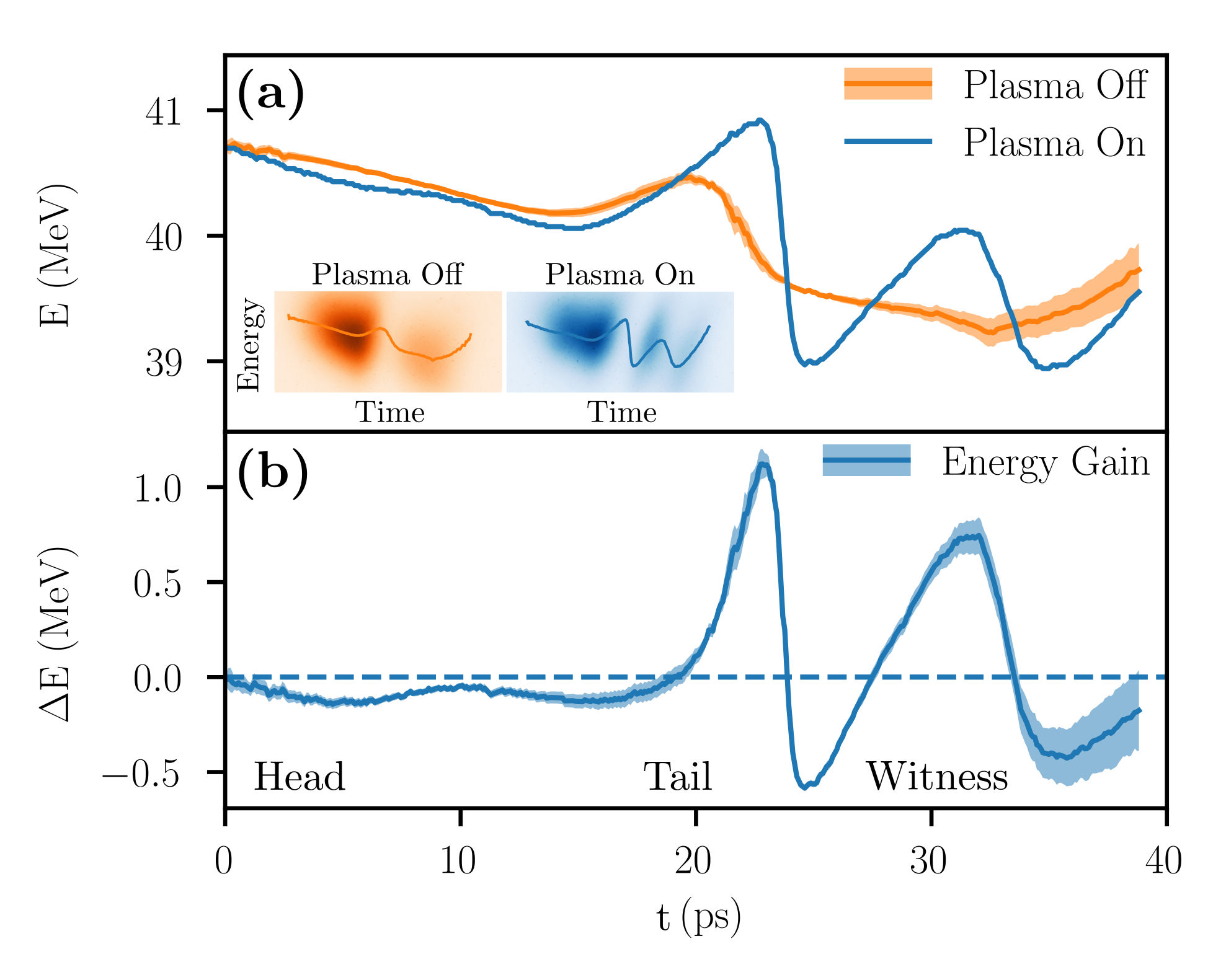}
	\caption{Example of the single shot wakefield measurement.
		(a) Averaged time slice energy centroid of 34 plasma off LPS measurements and a single plasma on LPS measurement. Inset: Raw YAG screen images of LPS diagnostic with individual time slice energy centroid for both plasma on and  off cases.
		Shading denotes 1 sigma variance in the background slice energy centroid.
		(b) Calculated difference between background and plasma on beam slice energy.}
	\label{fig:3}
\end{figure}

During the experiment, the beam was transported through the plasma source while  discharges were triggered at half the repetition rate of the accelerator  to acquire alternating plasma-on/plasma-off shots, eliminating systematic variations in the LPS present before the plasma. The measured time-slice energy centroids of the plasma-off shots were averaged as a background, to account for beam jitter and charge fluctuations. This background was then subtracted from single plasma-on measurements, to give the wakefield profile (Fig. \ref{fig:3}).
The unperturbed beam has a notable remaining longitudinal energy chirp from the EEX shaping process. The beam transported through plasma shows a high energy tail in the drive and a periodic modulation of the witness.

To quantify the beam current profile at the plasma interaction point, a reconstruction technique for calculating the longitudinal current density $\lambda_b(\xi)$ was developed that utilizes the longitudinal wakefield response $E(\xi)$ \cite{roussel_PRAB},
\begin{equation}
    \lambda_b(\xi)\propto\frac{\epsilon_0}{e}\Big(\frac{dE(\xi)}{d\xi} + k_p^2\int_{-\infty}^{\xi}E(\xi')d\xi'\Big),
\end{equation}
where $\xi = z - ct$ and $k_p$ is the wave number corresponding to the plasma density.
This reconstruction assumes a constant radial form factor, and considers the plasma wakefield response inside of the beam to be linear. Simulation studies have shown this to be an excellent approximation when the blowout region is small compared to the plasma wavelength \cite{lu_limits_2005,lu_nonlinear_2006}.
This method is used to reconstruct the time averaged drive current profile during the wakefield interaction from the measured energy changes.

\begin{figure}[!tbh]
    \includegraphics[width=\linewidth]{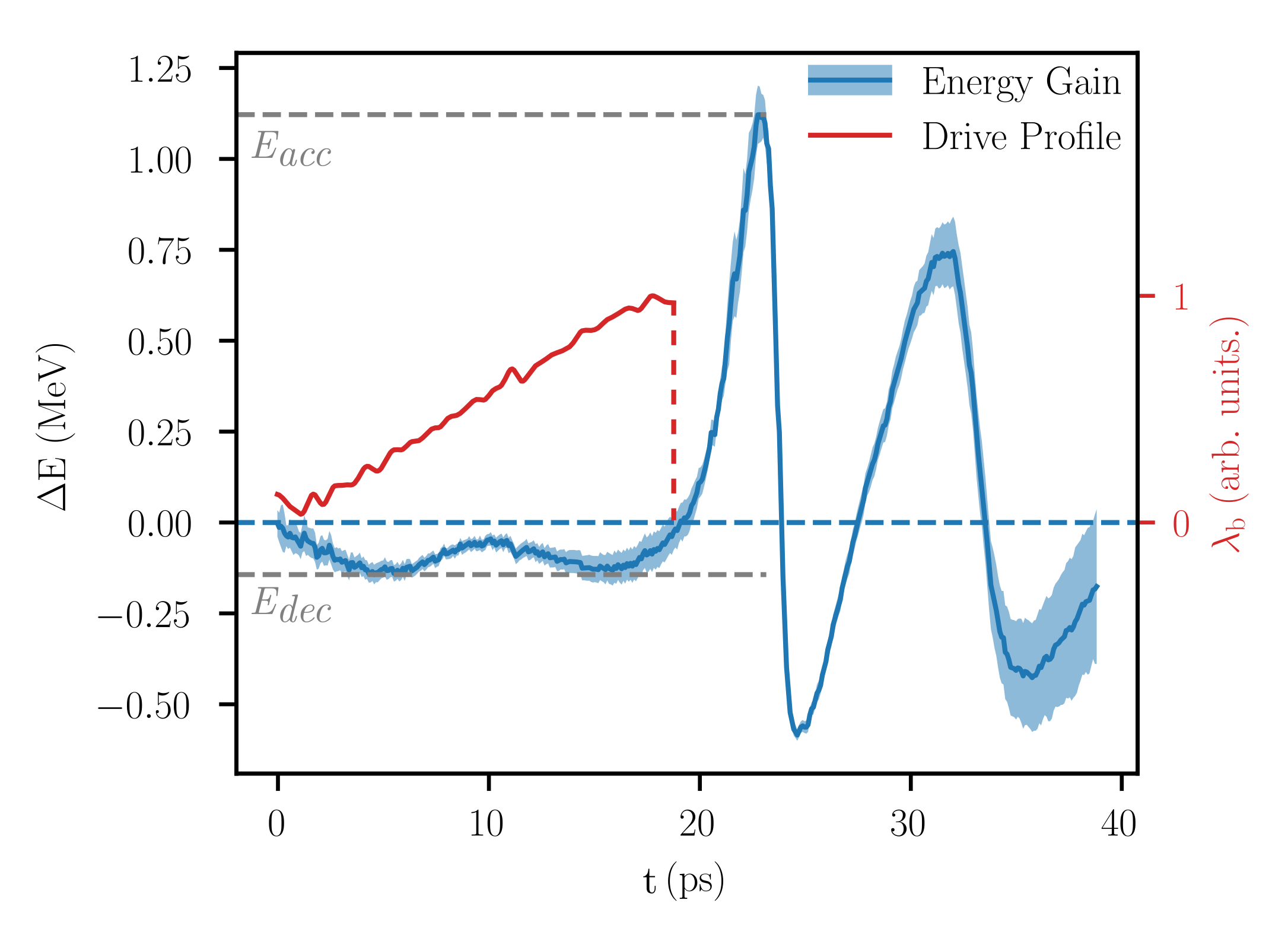}
    \caption{(Color online) Reconstructed current profile up to radial form factor showing linearly ramped driver (red) and associated energy difference (blue) from the averaged wakefield showing $\mathcal{R} = 7.8^{+2.5}_{-1.7}$.}
    \label{fig:4}
\end{figure}

The measured energy gain from the drive created from the mask of Fig. \ref{fig:1}b is shown in Fig. \ref{fig:4}, along with a reconstruction of the drive current profile.
The reconstructed drive profile shows a nearly linear ramp over two plasma wavelengths as evidenced by the number of undulations in the decelerating wake - the separation between neighboring minima of the wakefield in the drive is roughly equal to the plasma wavelength.
In the case shown, a separation of 10 ps corresponds to a plasma wavelength of 3 mm and thus an average plasma density of 1.3 x $10^{14}$ cm$^{-3}$.

In contrast to the drive region, the observed energy modulation of the witness shows clear evidence of a nonlinear plasma wakefield response.
As shown in simulation (Fig. \ref{fig:4}) the nonlinear rarefaction and subsequent collapse of plasma electrons to the beam axis produces a characteristic sawtooth longitudinal wakefield \cite{rosenzweig_acceleration_1991}. This wakefield form is evident in the witness bunch energy gain measurement. This observation also validates theoretical predictions that the maximum accelerating potential is asymmetric with respect to the maximum decelerating potential \cite{rosenzweig_trapping_1988}.
Furthermore, as the plasma wave in the nonlinear wake regime contains a number of harmonics \cite{rosenzweig_experimental_1989} the transformer ratio may be substantially increased compared to a single-mode wakefield. Indeed, the observed transformer ratio of $\mathcal{R} = 7.8^{+2.5}_{-1.7}$ matches well with this prediction, as it well-exceeds $\mathcal{R} < 6.3$ for the same profile in the linear, single-mode regime \cite{bane_collinear_1985}. This result is supported by the 3D PIC simulations. 

\begin{figure}[!tbh]
    \includegraphics[width=\linewidth]{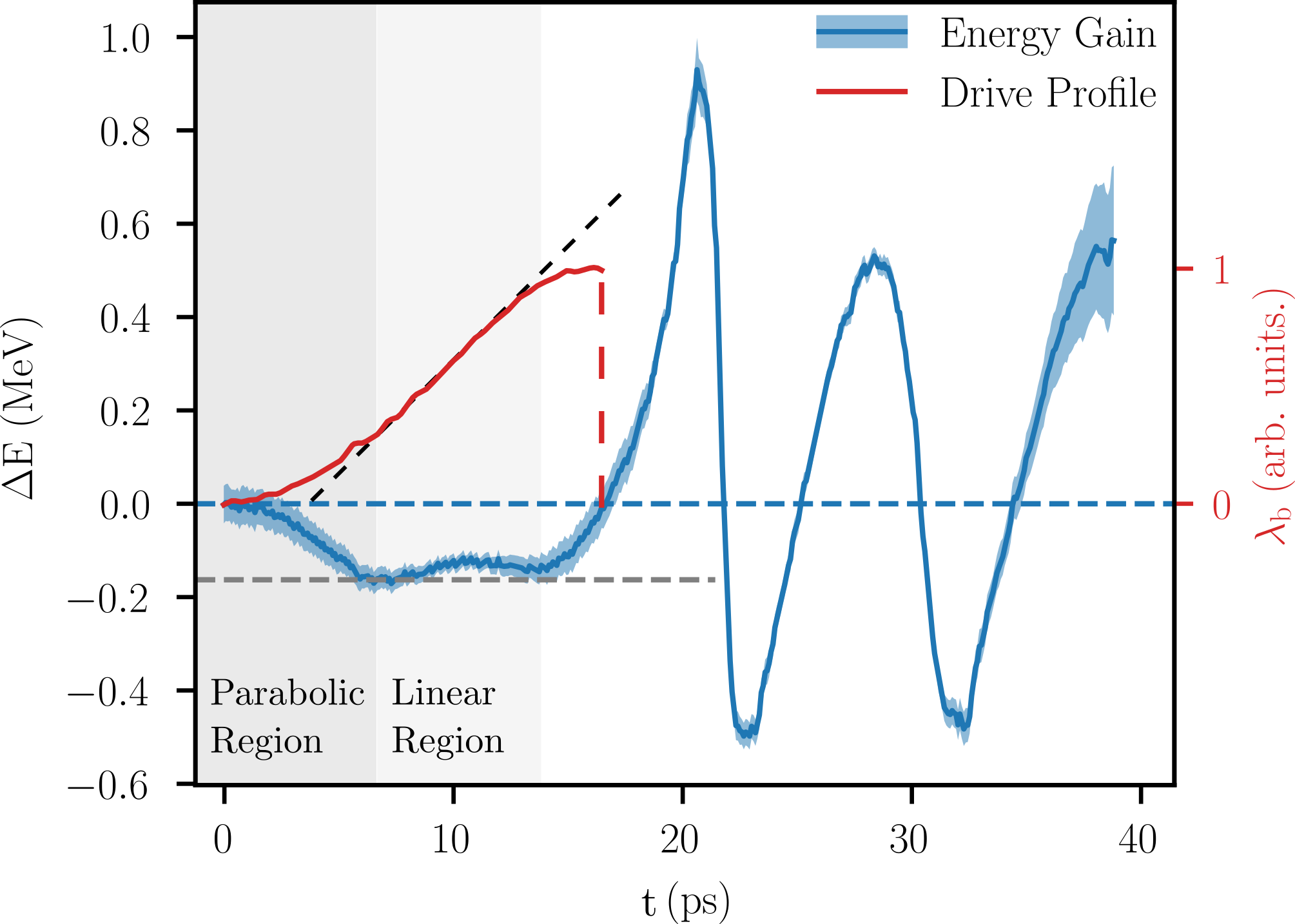}
    \caption{(Color online) Reconstructed current profile up to radial form factor showing linearly ramped driver with parabolic head (red), linear fit of reconstruction (black), and energy gain (blue) from the averaged wakefield showing a region of uniform drive deceleration.}
    \label{fig:5}
\end{figure}

While linearly ramped profiles provide a way to achieve high transformer ratios, perturbations to the head of the driver may provide further benefits in wakefield excitation. 
For example, the ramp can be modified through the addition of a shaped head, which acts to flatten the decelerating field inside the drive beam.
This is desirable, as it extends the beam-plasma interaction length by preventing partial depletion of the drive, while improving the overall transformer ratio.
Figure \ref{fig:5} shows a direct observation of this effect, in a case where the head of the reconstructed drive bunch has a parabolic shape, resulting in a near uniform deceleration of electrons in the linear region of the drive. 
The parabolic head has a length of 6.7 ps which approaches one plasma period (8.9 ps). 
The plasma wavelength was calculated by measuring the time between zero crossings in the witness ($n_0=1.6 $ x $10^{14}$ cm$^{-3}$) in this case, as opposed to inside the drive, due to the lack of distinct oscillations in the drive energy gain.
This observation shows good agreement with analytical calculations of the wakefield \cite{lemery_tailored_2015}, as uniform deceleration inside the drive occurs when the parabolic head length approaches an integer wavelength long. 

In summary, we demonstrate single-shot wakefield measurements of a PWFA in the nonlinear regime, with control over drive beam shaping through use of the EEX method.
The two key findings from the experiment are a recorded transformer ratio of $7.8^{+2.8}_{-1.8}$ from a linearly ramped beam, and the observation of a near uniform decelerating field for a drive beam with a parabolic head. 
The former is significant because enhancing the transformer ratio is a critical feature demanded in the design and deployment of next-generation wakefield accelerators \cite{none_advanced_2016}. The latter measurement complements and enhances this finding, because the uniform energy loss condition permits a longer sustained interaction, as one expects in an optimized transformer ratio case.
These aggregate results validate the importance of experimental efforts in longitudinal beam shaping. Further, they open the door to  accessing nonlinear PWFA conditions in a regime with less demanding requirements on the drive and plasma density than the use of an ultra-short bunch \cite{hogan_plasma_2010}. 
These advantages must be weighed against possible enhancement of beam breakup effects in long beams, which is a subject of future experimental investigations.
\begin{acknowledgments}
This work is supported by the Department of Energy, Office of High Energy Physics , under contract No. DE-SC0017648.
\end{acknowledgments}

%

\end{document}